\documentclass[twoside]{article}
\usepackage{fleqn,espcrc2}

\newcommand{\AmS}{{\protect\the\textfont2
  A\kern-.1667em\lower.5ex\hbox{M}\kern-.125emS}}

\def\beq{\begin{equation}} 
\def\eeq{\end{equation}} 
\def\bea{\begin{eqnarray}}  
\def\eea{\end{eqnarray}}
\def\bi{\begin{itemize}}  
\def\ei{\end{itemize}}  
\def\beqa{\begin{eqnarray}}  
\def\eeqa{\end{eqnarray}}

\def\pa{\partial}  
\def\cp{{\cal P}}  
\def\cl{{\cal L}}

\def\r2{\sqrt{2}}

\def\kaps{{\kappa}^{2}}  
\def\ov{\overline}  
\def\nn{\nonumber \\}

\def\lc{{\cal L}}  
\def\ca{{\cal A}}

\title{Supersymmetry and Vacuum Energy in Five-Dimensional Brane Worlds}

\author{Zygmunt Lalak\address{Theory Division, CERN\\CH-1211 
Geneva 23}\address{
Institute of Theoretical Physics\\
University of Warsaw}%
\thanks{This work has been supported by RTN programs HPRN-CT-2000-00152
and HPRN-CT-2000-00148 and   
by the Polish Committee for Scientific Research grant
5~P03B~119 20 (2001-2002).}}

\begin{document}
\thispagestyle{empty}

\begin{flushright}   CERN-TH/2001-068
\end{flushright}
\vskip 2cm
\begin{center}
{\huge Supersymmetry and Vacuum Energy in Five-Dimensional Brane Worlds}
\vspace*{5mm} \vspace*{1cm} 
\end{center}
\vspace*{5mm} \noindent
\vskip 0.5cm
\centerline{\bf Zygmunt Lalak${}^{1,2}$}
\vskip 1cm
\centerline{\em ${}^{1}$Theory Division, CERN}
\centerline{\em CH-1211 Geneva 23}
\vskip 0.3cm
\centerline{\em ${}^{2}$Institute of Theoretical Physics}
\centerline{\em University of Warsaw, Poland}
\vskip2cm

\centerline{\bf Abstract}
\vskip .3cm
We present the explicit form of the  four-dimensional 
effective supergravity action which describes low-energy physics 
of the Randall--Sundrum model with moduli fields 
in the bulk and charged chiral matter living on the branes.
The low-energy
 action is derived  from the compactification of a locally supersymmetric model in five dimension. 
We describe the mechanism of supersymmetry breaking mediation which relies 
on the non-trivial configuration of the  $Z_2$-odd  bulk fields.
Broken supersymmetry leads to stabilization of the interbrane distance.

\vskip1cm
\begin{flushleft}   
Talk given at SUSY2K, CERN, June 2000 and at 
``Thirty Years of Supersymmetry'', 
Minneapolis, October 2000.\\
%February  2001
\end{flushleft}
\newpage

\begin{abstract}
We present the explicit form of the  four-dimensional 
effective supergravity action, which describes low-energy physics 
of the Randall--Sundrum model with moduli fields 
in the bulk and charged chiral matter living on the branes.
The low-energy
 action is derived  from the compactification of a locally supersymmetric model in five dimensions. 
We describe the mechanism of supersymmetry breaking mediation which relies 
on the non-trivial configuration of the  $Z_2$-odd  bulk fields.
Broken supersymmetry leads to stabilization of the interbrane distance. 
\vspace{1pc}
\end{abstract}

\maketitle

\section{Introduction}
This talk is based on results obtained in collaboration with 
Adam Falkowski and Stefan Pokorski, published in refs. 
\cite{flp,flp2,flp3}. \\

Brane worlds with warped geometries offer new perspectives in 
understanding the hierarchy of mass scales in field theory models 
\cite{rs1,rs2,rs3}. The initial hope was that the mere presence of extra 
dimensions 
would be a natural tool to control mass scales in gauge theories 
coupled to gravity. However, the realization of these simple ideas 
in terms of consistent models eventually called for quite 
sophisticated constructions, such as the brane--bulk supersymmetry that 
we are going to discuss in this talk. \\
The basic five-dimensional setup of brane world models is that of four-dimensional hypersurfaces (branes) hosting familiar gauge and charged matter fields, 
which are embedded in a five-dimensional ambient space, the bulk, populated 
by gravitational and gauge-neutral fields. The bulk degrees of freedom 
couple to the fields living on branes through various types of interactions. 
Some of these interactions are analogues of an interaction between 
electromagnetic potential and charge density located on branes -- this is the case of the fields that are $Z_2$-even on an $S^1/Z_2$ orbifold forming the 
fifth dimension; some of them are rather analogues of the derivative coupling of the potential to the electric dipole moment density located on branes,
i.e. analogues of the interactions of the $Z_2$-odd fields on $S^1/Z_2$. 
These interactions lead to the formation of nontrivial vacuum 
configurations in the brane system. In particular, the solutions 
of Einstein equations of the form $ds^2 = a^{2}(x^5) ds^{2}_{4} + b^2(x^5) (dx^5)^2$ are usually allowed, as in the original Randall--Sundrum (RS)
models, where $ds^{2}_{4}$ is the Minkowski, anti-deSitter or deSitter 
metric in 4d. The two basic observations pertaining to the hierarchy problem
are the following. First, on the brane located at a position $x^5$, all the 
fundamental mass scales defining the 5d Lagrangian become down-scaled 
by the factor $a(x^5)$: $m \rightarrow m a(x^5)$ when written down in the frame canonical with respect to the 4d line element $ds^{2}_4$. Thus, if the 
warp factor falls down exponentially, as in the RS model, one is given a 
natural exponential mass hierarchy between branes which is directly related 
to their separation. In fact, the effective  mass measuring 
the interaction of a test body with the gravitational zero-mode is 
modulated by the warp factor, $m_{eff} = m a(x^5)$. In addition, the 
heavy Kaluza-Klein modes of the metric tensor couple to the brane matter 
energy momentum tensor at $x^5$ with the strength $\Lambda^{-1} (x^5) $ 
where $ \Lambda (x^5) = M_P a(x^5)$, thus implying, at least naively, 
an $UV$  cut-off of that scale on perturbative physics on the brane. 
Second, as pointed out long ago by Rubakov and Shaposhnikov \cite{RuSha},
the gradient energy associated with the variation of the warp factor in the direction transverse to the brane can cancel the contribution of the brane 
physics to the effective 4d cosmological constant. However, the     
stumbling observation is that whenever one finds a flat 4d foliation 
as the solution of higher-dimensional Einstein equations, which seems to be 
necessary for the existence of a realistic 4d effective theory, 
it is accompanied by a special choice of various parameters in the higher 
dimensional Lagrangian (see \cite{flln}). The fine-tuning seems to be even worse in 5d than 
in 4d, since typically one must correlate parameters living on spatially 
separated branes. Then there appears immediately the problem of stabilizing 
these special relations against  quantum corrections. 
This situation has prompted the proposal \cite{bagger,gp,flp}, that it 
is 
a version of brane-bulk supersymmetry that may be able to explain 
apparent fine-tunings and stabilize hierarchies against quantum corrections.
Indeed, the brane-bulk supersymmetry turns out to correlate in the right 
way the brane tensions and bulk cosmological constant in the supersymmetric Randall--Sundrum model. 
Moreover, local supersymmetry is likely to be necessary to embed brane worlds 
in string theory. 
Hence, the quest for consistent supersymmetric 
versions of
brane worlds goes on, see \cite{bagger,gp,flp,hiszp,flp2,kallosh,bc,bc2,kl,kl2,kb,pmayr,duff,louis,zucker,bagger2,ls,flp3,clp}. 
First  explicit supersymmetric 
models with 
delta-type (thin) branes were constructed in \cite{bagger,flp,flp2,kallosh}. 
The distinguishing feature of the pure supergravity Lagrangians 
proposed in \cite{flp} is imposing the $Z_2$ 
symmetry,
such that gravitino masses are $Z_2$-odd. An elegant formulation of the 
model is given in ref. \cite{kallosh} where an additional non propagating fields are introduced to independently supersymmetrize the branes 
and the bulk. In the on-shell picture for these  fields the models 
of ref. \cite{kallosh} and refs. \cite{flp,flp2} are the same. 
On the other hand, in refs. \cite{flp,flp2} 
it has 
been noted, that supersymmetric Randall--Sundrum-type models can be generalized 
to include the universal hypermultiplet and gauge fields and matter on the 
branes. The 
Lagrangian of such a construction has been given in \cite{flp,flp2,flp3,mt}.
This has allowed us \cite{flp,flp3} to study issues such as supersymmetry 
breaking and its transmission through the bulk. 
Finally, in \cite{flp3} the effective low energy 
theory was formulated, which describes properly  the physics of the warped 
five dimensional models with  gauge sectors on the branes. 

We have shown that in the class of 
models without nontrivial gauge sectors in the bulk, 
unbroken $N=1$ local supersymmetry (classical solutions with four unbroken 
supercharges) implies vanishing of the effective cosmological 
constant.
We have demonstrated  the link between vanishing of the 4d cosmological 
constant, minimization of effective potentials in 5d and 4d, and moduli 
stabilization.
We have also described supersymmetry breaking due to a global obstruction
against the extension of bulk Killing spinors to the branes, which is a
phenomenon observed earlier in the Horava--Witten model in 11d and 5d.

First steps towards the 4d effective theory were made in \cite{bagger2},\cite{ls}
(where the K\"ahler function for the radion field was identified).
In the set-up we consider in this paper supersymmetry 
in 5d is first broken from eight down to four supercharges by the BPS 
vacuum wall, and then again broken spontanously down to $N=0$ due to 
switching on expectation values of sources living on the branes. 
The general strategy follows the one 
\cite{peskin-mirabelli,ovrutdw,ovrut,elpp} 
that led to the complete and accurate description of the low-energy 
supersymmetry breakdown in the Horava--Witten models, see
 \cite{peskin-mirabelli,elpp,elp}. We were able to deduce 
the K\"ahler potential, 
superpotential and gauge kinetic functions describing physics of corresponding vacua in four dimensions. It turns out that the warped background modifies 
in an interesting way the kinetic terms for matter fields and the gauge kinetic function on the warped wall. There also appears a potential for the radion superfield, its origin being a modulus-dependent prefactor multiplying 
the superpotential on the warped wall in the expression for the 
4d effective superpotential. 
%We do not need to introduce any nontrivial 
%gauge sector in the bulk to generate a potential for the $T$ modulus. It is interesting to note the structure of the effective 4d supergravity is completely different than that of the no-scale models 
%\cite{ns1,ns2}
%. In the no-scale models the 4d cosmological constant vanishes, while $F_T$ is undetermined and sets the supersymmetry breaking scale. In our model $F_T$ vanishes, supersymmetry is broken by $F_S$ and non-zero cosmological constant is induced. 
%The complete and phenomenologically relevant 4d N=1 supergravity 
%model which we managed to construct in this paper should finally 
%facilitate 
%a detailed investigation of the low energy physics of 
%warped compactifications.     

\section{Supersymmetry in the brane-world scenarios}
Let us begin with a brief review of the original RS model. The action is that of   
 5d gravity on $M_4 \times S_1/Z_2$, with  negative cosmological constant:
\bea
\label{rsaction}
&S=M^3 \int d^5x \sqrt{-g}(\frac{1}{2}R+6 k^2)& \nn 
&+ \int d^5x \sqrt{-g_{i}} (-\lambda_1 \delta(x^5)- \lambda_2\delta(x^5-\pi\rho) ). &
\eea 
Three-branes  of non-zero tension are located at $Z_2$ fixed points.   The  ansatz for vacuum solution preserving 4d Poincare invariance has the warped product form:
\beq 
\label{rsansatz}
ds^2= a^2(x^5)\eta_{\mu\nu}dx^\mu dx^\nu+R_0^2(dx^5)^2.
 \eeq
The breathing mode of the  fifth dimension is parametrized by $R_0$.
The solution for the warp factor $a(x^5)$ is: 
\beq
\label{exp}
a(x^5)=\exp(-R_0 k |x^5|).
\eeq
It  has an exponential form, which can generate large hierarchy of scales between the branes. Matching delta functions in the equations of motion requires fine-tuning  of the brane tensions:
\beq
\label{ft}
\lambda_1=-\lambda_2=6k.
\eeq 
With the choice (\ref{ft}) the matching conditions are satisfied for arbitrary $R_0$, so the fifth dimension is not stabilized in the original RS model. Thus $R_0$ enters the 4d effective theory as a massless scalar (radion),
 which couples to gravity in the manner of a Brans--Dicke scalar.  This is at odds with the precision tests of general relativity, so any realistic model should contain a potential for the radion field.    

Relaxing the condition (\ref{ft}) we are still able to find a solution in the maximally symmetric form, but only if we allow for  non-zero 4d curvature ($adS_4$ or $dS_4$) \cite{rk}. In such a case the radion is stabilized and its vacuum expectation value is determined by the brane tensions and the bulk cosmological constant.  

The Randall--Sundrum model can be extended to a locally supersymmetric model \cite{bagger,gp,flp}. The basic set-up consists of 5d $N=2$  gauged supergravity \cite{gunaydin,agata}, 
which includes the gravity multiplet $(e_\alpha^m, \psi_\alpha^A, \ca_\alpha)$, that is the metric (vielbein), a pair of symplectic Majorana gravitinos, and a vector field called the graviphoton. The 5d SUGRA 
action is 
\beqa 
%\label{5daction} 
&S=\int d^5xe_{5}\frac{1}{\kaps} \left ( \right .  
\frac{1}{2}R -\frac{1}{2}\ov{\psi_\alpha}^A\gamma^{\alpha\beta\gamma}D_\beta\psi_{A\gamma}& \nn
&-\frac{3}{4}{\cal F}_{\alpha\beta}{\cal F}^{\alpha\beta}+ ...
   \left . \right)  & 
\eeqa 
and the supersymmetry transformations are given by 
\beqa 
\label{zuzanna}
&\delta e_{\alpha}^{m}= 
\frac{1}{2}\ov{\epsilon}^A\gamma^{m}\psi_{A\alpha}& 
 \nonumber \\ 
&\delta \psi_\alpha^{A}= 
D_{\alpha}\epsilon^{A} 
-\frac{i}{4\sqrt{2}}(\gamma_{\alpha}^{\;\beta\gamma}-4\delta_{\alpha}^{\beta}\gamma^{\gamma}){\cal F}_{\beta\gamma}\epsilon^{A}
& 
\nonumber \\ 
&\delta{\cal A}_{\alpha}= 
-\frac{i}{2\sqrt{2}}\ov{\psi_\alpha}^A\epsilon_A .&  
\nonumber
\eeqa
Let us now add the brane tension at the brane located at $x^5=0$, $S_1= 
\int d^5x e_4 (-6 k) \delta(x^5)$, and perform the supersymmetry 
transformation on the determinant of the induced vierbein. This produces 
a delta-type variation in the action:   
$\delta e \Rightarrow 3 \delta(x^5) e_4  k ( \bar{\psi^1}_\mu \gamma^\mu \epsilon^1 + (1 \leftrightarrow 2))$. It is straightforward to notice that this can be cancelled 
through the variation of the term $\bar{\psi^A}_\mu \gamma^{\mu 5 \rho} D_5 
\psi^{A}_\rho$ upon introducing new terms in the transformations of gravitini:
$\delta \psi^{1}_\alpha = + \frac{ k}{2} \epsilon(x^5) \gamma_\alpha 
\epsilon^1$,
$ \delta \psi^{2}_\alpha = - \frac{k}{2} \epsilon(x^5) \gamma_\alpha 
\epsilon^2$. These corrections introduce further variations in the bulk Lagrangian, which require further new terms in the bulk Lagrangian:
${\cal L}_{\psi^2} = + \frac{3 e_5}{4} k \epsilon(x^5) ( \bar{\psi^1}_\alpha \gamma^{\alpha \beta} \psi^{1}_{\beta} - \bar{\psi^2}_\alpha \gamma^{\alpha \beta} \psi^{2}_{\beta})$ 
and 
${\cal L}_{cc} = 6 e_5 k^2 $, which is precisely the bulk potential needed in 
the RS model. The
continuation through $x^5 = \pi \rho$ gives on the second brane the tension 
term 
$+ \delta (x^5 - \pi \rho) e_4 6 k$. 
The resulting locally supersymmetric Lagrangian is in fact that of a 
gauged supergravity. The symmetry that is gauged is the
$U(1)$ subgroup of the R-symmetry, $\psi_\alpha^A \rightarrow e^{i\phi}\psi_\alpha^A$, the gauge field being 
$\ca_\alpha^R=-\frac{1}{2\r2}\ca_\alpha$.
Gauging of the $U(1)_R$ symmetry means 
that we replace the derivative acting on the gravitino with the $U(1)_R$ covariant derivative:
\beqa
&D_\alpha \psi_\beta^A \rightarrow D_\alpha \psi_\beta^A 
-\frac{3}{\r2}(\sigma^3)^A_{\;B} k \epsilon (x^5)\ca_\alpha \psi_\beta^B,
& \nonumber
\eeqa 
 where $D_\alpha$ denotes the ordinary space-time covariant derivative.
The coefficient of the coupling $\ca_\alpha \psi_\beta^B$ defines 
the prepotential $ {\cal P} = \frac{i }{4} \sigma^3$
 and the $Z_2$-odd gauge coupling $g= \frac{6 k 
\epsilon(x^5) }{\sqrt{2}}$.
The warp factor turns out to be  $a(x^5) = e^{-k R_0 |x^5|}$, precisely the one of the original RS model.
Thus the fine-tuning present in the original RS model can be explained by the requirement of local  supersymmetry \cite{flp}.

 New bosonic and fermionic fields do not affect the vacuum solution,
so that the equations of motion  for the warp factor are the same as in the original, non-supersymmetric RS model. 
The RS solution  satisfies the BPS conditions and preserves one half of the supercharges, which corresponds to unbroken $N=1$ supersymmetry 
in four dimensions.   
 In the supersymmetric version the brane tensions are fixed. In consequence, the exponential solution (\ref{exp}) is the only maximally symmetric solution and the radion is still not stabilized.

We now turn to studying the supersymmetric RS model coupled to matter fields.  We want to investigate how general the features present in the minimal supersymmetric RS model are.  We find that unbroken local supersymmetry implies flat 4d space-time   in a wider class of 5d supergravities coupled to hyper- or vector multiplets, in which the scalar potential is generated by gauging a subgroup of R-symmetry.

Let us add neutral matter in the bulk (moduli) in the form of a universal 
hypermultiplet, $(\lambda^a, V, \sigma, \xi, \bar{\xi})$.
The quaternionic metric $h_{uv}$ of the scalar manifold 
can be read from the K\"ahler potential: 
$ K=-\ln(S+\bar{S}-2\xi\bar{\xi}), \;\;
S=V+\xi\bar{\xi}+i\sigma. $ \\
New kinetic terms in the bulk Lagrangian contain 
\beqa 
\label{5dactionhip} 
&S=\int d^5xe_{5}\frac{1}{\kaps} \left ( \right .   
-\frac{1}{4V^2}(\pa_{\alpha}V\pa^{\alpha}V+
\pa_{\alpha}\sigma \pa^{\alpha}\sigma) 
& \nonumber \\
&-\frac {1}{V}\pa_{\alpha}\xi\pa^{\alpha}\bar{\xi}
  -\frac{1}{2}\ov{\lambda}^a\gamma^\alpha D_\alpha \lambda_a  + ...\left . \right).  & \eeqa 
%The Sp(1) indices on the hyperini are raised and lowered with $\Omega^{ab}$.
%We choose $\Omega^{21}=\Omega_{21}=1$.
Under $Z_2$, the  bosonic fields 
$(V, \sigma)$ are even, and $(\xi)$ is odd. 
To arrive at supersymmetric Lagrangian in the presence of these additional 
fields it is necessary to gauge, in addition to $U(1)_R$, the isometry of the hypermultiplet scalar 
manifold: $\xi \rightarrow e^{i\theta} \xi$. Components of the Killing vector corresponding to this isometry are $
k^\xi = i \xi, \;\; k^{\bar{\xi}} = -i \bar{\xi},\; \;\;\; k^x=-y, \;\; k^y=x 
$, where $\xi = x + i y$. This results in the prepotential 
\beqa 
\label{ourprepotential} &
\cp= 
(\frac{1}{4}- \frac{x^2+y^2}{4V}) i \sigma^3 - \frac{x}{2V^{1/2}} i\sigma^1 
+ \frac{y}{2V^{1/2}} i\sigma^2 & 
\eeqa
and scalar potential:
\beqa &
V = - k^2 \left ( 6
 +\frac{3}{ V}|\xi|^2 - \frac{3}{V^2}|\xi|^4 \right).& 
\eeqa   
Finally, supersymmetry requires new terms localized on  the branes:
$\lc_B=-\frac{e_4}{\kaps} 6 k  (1-\frac{|\xi|^2}{V}) (\delta(x^5) - \delta(x^5-\pi\rho))$ and modifications of supersymmetry transformation laws that were given in \cite{flp3}. 
To discuss both unbroken and broken $N=1$ supersymmetry, let us use 
from now on a more general ansatz for the metric 
allowing $adS_4$ geometry in four dimensions (the case of $dS_4$ is similar):
$g_{\alpha \beta} = \; diag \, (-a^2 (x^5 ) e^{2 L x^3}, a^2 (x^5 ) e^{2 L x^3}, a^2 (x^5 ) e^{2 L x^3},\\ a^2 (x^5), R_0^2),$ 
where $R_0$ is the radion, and for later convenience we define $f=\exp(L x^3)$.
Let us note that
$R^{(4)}=-12 L^2$ and $\Lambda^{(4)}_{cc} = \langle V_{eff} \rangle / M^{2}_P=  -6 L^2$.
The conditions for unbroken supersymmetry in the bulk (BPS conditions)
$\delta \psi_\alpha^A = \delta \lambda^a= 0$
correlate scalars, the warp factor, Killing vector and 
prepotential:
\bea
&\frac{a'}{a}=-4 k  R_0\sqrt{\vec{P}^2}, \;
h_{uw}\pa_5 q^w =6 k R_0 \pa_u \sqrt{\vec{P}^2}& 
\nonumber \\
&h_{uv}\pa_5 q^u\pa_5 q^v= 9 R_0^2 k^2 h_{uv} k^u k^v.&
\eea 
After substituting these relations into the expression for the 4d action and 
integrating over $x^5$ one obtains 4d effective action of the form
\bea 
&
S_4=M_{PL}^2\int d^4x (\frac{1}{2}\bar{R}-V_{eff})
&\nn
&M_{PL}^2 V_{eff}=- M^3 a^4(0)(24 k \sqrt{\vec{P}^2}(0)-\lambda_1)& \nn
&+ M^3 a^4(\pi\rho)(-24 k \sqrt{\vec{P}^2}(\pi\rho)-\lambda_2).& 
\eea
Unbroken supersymmetry corresponds to $\sqrt{\vec{P}^2}=P_3$, 
which immediately 
implies $V_{eff}=\langle R^{(4)} \rangle = 0$.
This is a general result, which holds for any hypermultiplet manifold, not only for the universal one, which implies that 
unbroken SUSY excludes solutions with non-zero 4d space-time curvature. 
Only Minkowski foliations correspond to unbroken $N=1$ supersymmetry
and the converse is also true in the class of models which are discussed in 
ref. \cite{flp3}.  
A further consequence is that stabilization of the radion field is impossible without breaking of residual four supercharges. It should be noted that      
in 4d supergravities one can a'priori obtain an anti-deSitter solution and  preserve supersymmetry at the same time. Hence, 
compactifications of the supersymmetric RS scenarios yield a special 
subclass of 4d supergravities. 

\section{Supersymmetry breaking mediated by $Z_2$-odd fields and radion 
stabilization}
In this section we investigate an alternative mechanism of supersymmetry breaking, similar to that studied in M-theoretical scenarios \cite{elpp}. 
It is triggered by brane sources coupled to the scalar fields in the bulk, which are odd with respect to the $Z_2$ parity. One way to see that supersymmetry is broken is to notice that the Killing spinor cannot be defined globally. The odd fields are the agents that transmit supersymmetry breaking between the hidden and visible branes. We present below a general description of our mechanism and then apply it to a specific model of 5d gauged supergravity with the universal hypermultiplet. 

To see how to break the remaining supersymmetry (corresponding to $N=1$ in 4d) 
down to nothing ($N=0$ in 4d) one should inspect the parts of supersymmetry transformations of bulk fermions that contain the $Z_2$-odd fields $\xi$ and their transverse derivatives: 
\bea
&\delta \psi^{1}_{5}  = -\frac{1}{\sqrt{V}} \pa_5 \xi \epsilon^2 
+ 2 k \epsilon(x^5)
 (-\frac{Re(\xi)}{2 \sqrt{V}} (\sigma^{1})^{1 B}& \nn
&+\frac{Im(\xi)}{2 \sqrt{V}} (\sigma^{2})^{1 B}) 
 \gamma_5 \epsilon_B, & \nonumber \\
&\delta \lambda^1= +\frac{i}{\sqrt{2V}} \gamma^5 \pa_5 \xi \epsilon^2
+ 3 k \epsilon(x^5)
V^{A1}_u k^u  \epsilon_A.& 
\eea
It is clear that 
vacuum expectation values of $\xi$ and $\pa_5 \xi$ are parameters of the 
supersymmetry breakdown and the field $\xi$ is the agent of local mediation
of supersymmetry breakdown. 
Another way to see that $\xi, \; \pa_5 \xi \neq 0$ imply  completely broken 
SUSY is to notice that this means that the components ${\cal P}^1$ and 
${\cal P}^2$ of the prepotential are non-zero as well. The 
BPS conditions give:        
\beq
\epsilon^2=\frac{\sqrt{\vec{P}^2}\gamma_5\epsilon^1-P_3\epsilon^1}{P_1-iP_2}
. 
\eeq
As an immediate consequence we find that if $P_1$ or $P_2$ $\neq \;\;0$, then 
the Killing spinor is non-chiral, and is a mixture of the even and the odd components of the SUSY parameter $\epsilon$. 
But only the even component survives the $Z_2$ projection at the orbifold fixed points, hence the Killing spinor cannot be defined globally and
supersymmetry is broken because of the `misalignment' between the bulk and the brane supersymmetries. In other words, the projection of the general bulk 
Killing spinor onto the brane would contain insufficient degrees of freedom 
to generate the minimal supersymmetry on the brane.   
Explicitly, bulk BPS conditions require 
\beq
\label{hBPS}
\frac{a'}{a}=-4 k  R_0\sqrt{\vec{P}^2}, \;
h_{uw}\pa_5 q^w =6 k R_0 \pa_u \sqrt{\vec{P}^2}, 
\eeq 
but matching delta functions in the 
 equations of motion implies 
\bea 
&\frac{a'}{a}(0)=-4 k R_0 P_3(0),\;  
\frac{a'}{a}(\pi\rho)=-4 k R_0 P_3(\pi\rho)& \nn
&h_{uw}\pa_5 q^w(0)=6 k R_0 \pa_u P_3(0)& \nn
&h_{uw}\pa_5 q^w(\pi\rho)=6 k R_0 \pa_u P_3(\pi\rho).& 
\eea
All these conditions are automatically satisfied if $\sqrt{\vec{P}^2}=P_3$.
Thus
 $P \sim \sigma^3$ is equivalent to the RS fine-tuning, unbroken supersymmetry and 4d Poincare invariance.
However, if ${\cal P}_1$ or ${\cal P}_2$ are non-zero, there are 
four matching conditions, but only three free parameters;
%two integration constants ($a(0)$ drops out of the matching conditions) and th%e size of the fifth dimension $R_0$. 
generically we are not able to satisfy the boundary conditions,
and thus supersymmetry is broken. 

To excite $\xi,\; \pa_5 \xi$ one needs to couple them to the branes. 
Details are given in \cite{flp3}. Forgetting for a while about 
coupling to gaugino condensates on branes, the relevant part of the 
brane--bulk coupling is 
\bea
& -\int d^5 x \; e_5 \frac{2}{V g_{55}} ( 
\delta(x^5)W_1(\pa_5 \xi + 2 \delta(x^5) \bar{W}_1)& \nn    
&+ \delta(x^5-\pi\rho)
W_2 ( \pa_5 \xi +2 \delta(x^5-\pi\rho) \bar{W}_2)+ h.c. ).& \nn
& & 
\eea
We solve equations of motion perturbatively to order $(\frac{
W}{M^3})^2$  with $adS_4$ foliation and ansatz
$\xi=\xi(x^5)=\epsilon(x^5) \zeta(|x^5|), \; V=V(x^5), \;\sigma=const, 
\; \ca_\mu=0, \; \ca_5=const$.
Matching the $\delta'$ in EOMs yields the boundary conditions: $\zeta(0)=-\frac{\ov{W}_1}{M^3}, \;\;\zeta(\pi\rho)=\frac{\ov{W}_2}{M_3}$. 
As a consequence:
\bea
\label{bc:w}
&C= -\frac{\ov{W}_1}{M^3},& Ce^{k( R_0 -  3\sqrt{2}i\ca_5)\pi\rho}=
\frac{\ov{W}_2}{M^3}. 
\eea
One can see that as long as supersymmetry is unbroken, 
moduli $R_0$ and $\ca_5$ 
are arbitrary. When sources are switched on for odd fields, then 
the expectation value of the radion is determined by the boundary 
sources $W_i$, assuming that $V$ gets frozen.
Perturbative ($o(|W^2|)$) solution to EOMs is
\bea
\label{solutionall}
&\xi = C \epsilon(x^5) e^{k( R_0 -  3\sqrt{2}ik\ca_5)|y|},\; 
C= -\frac{\ov{W}_1}{M^3}, & \nn
& Ce^{k( R_0 -  3\sqrt{2}ik\ca_5)\pi\rho}=\frac{\ov{W}_2}{M^3}&\nn
&a=e^{kR_0|y|} + \frac{|C|^2}{2V}e^{-kR_0|y|},\;
 L^2=\frac{16}{6}\frac{k^2|C|^2}{V}&\nn
&V=V_0 -|C|^2 e^{2kR_0 |y|},\; \sigma=\sigma_0, & 
\eea
except that boundary conditions for $V$ are fullfilled at the zeroth order 
only; hence it is better to say that one assumes $V$ to be `frozen'. 
The four-dimensional curvature is $R^{(4)} = - 32 \frac{k^2 |C|^2}{V}$.
This means that in our family of models broken supersymmetry implies negative 
4d curvature, and that curvature vanishes only if the SUSY-breaking sources 
are switched off.  

As a cross-check of the above results we can calculate the 4d effective potential, obtained by integrating out the 5d bosonic action in the background (\ref{solutionall}). The result (to the order $\frac{W}{M^3}^2$) is:
\beq
\cl_4 = \sqrt{-\bar{g}}\frac{M^3}{k}(1-e^{-2kR_0\pi\rho}) \left (
\frac{1}{2}\bar{R} + 8\frac{k^2|C|^2}{V_0}\right).
\eeq
We denoted by $\bar{g}$ the oscillations of the 4d metric around the vacuum solution. Solving the Einstein equations in the 4d effective theory yields $\bar{R}=-32\frac{k^2|C|^2}{V_0}$, which is consistent with the value of $L^2 \equiv -\frac{1}{12}\bar{R}$ in (\ref{solutionall}). We also see that $V_0$ enters the denominator of the effective potential, which explains its runaway behaviour commented on earlier.
      
Before closing the discussion of the 5d classical solutions 
and supersymmetry breakdown, let us comment on proposals 
\cite{pmayr,verlinde,schmid} to solve the 
cosmological constant problem by virtue of 
supersymmetry of the bulk-brane system. 
To put the issue into perspective, let us note that the Einstein equation
with indices $(55)$ does not contain second 
derivatives of fields, hence it acts as a sort of constraint on the 
solutions of the remaining equations. This becomes clearer 
in the Hamiltonian 
approach towards the flow along the fifth dimension, where this equation 
arises as the Hamiltonian constraint ${\cal H} =0$, and is usually 
used to illustrate the way the conservation of the 4d curvature $ L^2$ 
is achieved through the compensation between gradient and potential terms 
along the classical flow. 
However, this classical conservation hinges upon 
fulfilling certain consistency conditions between brane sources, or 
between boundary conditions induced by them, as illustrated by the model 
above. When one perturbs the boundary terms on one wall, then 
to stay within the family of maximally symmetric foliations one of two things 
must happen. Either the distance between branes must change, or the 
source at the distant brane must be retuned. In the class of models 
which we constructed, if the 4d curvature is present then supersymmetry is broken, and doesn't take care of such a retuning. Moreover, even if retuning takes place,
the size of 4d curvature, i.e. of the effective cosmological constant,
does change as well, moreover, the magnitude of the effective cosmological 
constant has quadratic dependence on the boundary terms 
which induce supersymmetry breakdown. Hence, any perturbation of the 
boundary, instead of being screened by the bulk physics, contributes  
quadratically to the effective cosmological constant. 
Of course, we are talking about perturbations which can be considered 
quasi-classical on the brane.    
Thus we do not see here any special new effect of the extra dimension 
in the the cancellation of the cosmological constant. The positive 
aspect of supersymmetry is exactly the one which we know from the 4d physics.
Supersymmetry, even the broken one, limits the size of the brane terms 
inducing supersymmetry breakdown, and limits in this way the magnitude 
of the 4d cosmological constant, since the two effects are strictly related 
to each other. 

\section{Effective low-energy theory in four dimensions}
In this section we give the form of the effective four-dimensional 
 supergravity describing zero-mode fluctuations in the models presented in the previous sections. 
Since in the 5d set-up supersymmetry 
is broken spontaneously, it is safe to  assume 
that in 4d this supersymmetry breakdown can be considered as a spontaneous 
breakdown in a 4d supergravity Lagrangian described with the help of 
a K\"ahler potential $K$, superpotential $W$ and gauge kinetic functions
$H$. 
The goal is to identify these functions reliably starting from the 
maximally symmetric approximate solutions (\ref{solutionall}) that 
we have described in the previous section.  
Our procedure is perturbative in the supersymmetry breaking parameter $\frac{W}{M^3}$. Fortunately, 
it is  sufficient 
to identify the functions we are looking for from the terms that
can be reliably read at the order $(\frac{W_1}{M^3})^1$. Such terms include 
the gravitino mass term. In addition, we have 
at our disposal the complete kinetic terms for moduli, gauge and matter fields,
which are of order $(W_1)^0$ and are sufficient to read off the K\"ahler 
potential for moduli and matter fields. The complete procedure has been given 
in ref. \cite{flp3}. Here we summarize the results and discuss the basic features of the warped 4d supergravities. 

The K\"ahler function for moduli fields $S=V_0 + i \sigma_0$ and 
 $T= k\pi\rho (R_0 + i\sqrt{2}\ca_5)$, and for the charged fields $\Phi_1$ and 
$\Phi_2$ living on the Planck brane and warped brane respectively is  
\bea
&K(S,\bar{S};T,\bar{T};\Phi,\bar{\Phi})=-M^{2}_P \log(S + \bar{S})& \nn
& -3 M^{2}_P \log ( f(T+ \bar{T}
-\frac{k}{3 M^3} |\Phi_2|^2) -\frac{\beta k}{3 M^3} |\Phi_1|^2)&\nn
& &
\eea
where we defined $M_P^2= \frac{M^3}{k}(1-e^{-2k\pi\rho \langle R_0 \rangle })$, $f= \beta (1- e^{-(T + \bar{T})})$ with 
$\beta = \frac{ M^3}{k M^{2}_{P}}$.
The effective 4d superpotential is 
\bea
&W= 2 \sqrt{2} (W_1 + e^{-3 T} W_2 )& 
\eea
and the gauge kinetic functions are 
\beq 
H_{warped}(S,T) = S + 2 b_0 T, \; H_{Planck}(S)=S. \eeq

One can study the 4d effective scalar potential derived from $K,\;W,\;H$. 
The standard formulation gives: 
${\cal V} = e_4 e^{G}(G_iG^{i\bar{j}}G_{\bar{j}}-3 )$ with $G=K + 
\log |W|^2$. Explicitly:
\bea
&{\cal V}= e_4 \frac{4}{V_0 M^{2}_P \beta^3 (1-e^{-2k\pi\rho R_0})^3} (
|W_1|^2&\nn
&(3 e^{-2k\pi\rho R_0}-2) 
+ |W_2|^2 (3 e^{-4k\pi\rho R_0} - 2 e^{-6k\pi\rho R_0})
&\nn
&+W_1\bar{W_2}e^{-3k\pi\rho{R_0-i\r2\ca_5}}&\nn
&+W_2\bar{W_1}e^{-3k\pi\rho{R_0+i\r2\ca_5}}).& 
\eea
Minimizing the above scalar potential wrt $\ca_5$ yields 
\beqa 
&Arg(W_2)-Arg(W_1) -3 \sqrt{2} k \pi \rho {\cal A}_5 = \pi\, n,&\nn
&n=0,\pm 1,\pm2,...\;.& 
\eeqa
Taking $n=1$ and minimizing wrt $R_0$ one obtains 
\beqa &
e^{-k \pi \rho R_0} = \frac{|W_1|}{|W_2|},
& \eeqa
consistently with the 5d picture. 
Let us summarize the basic features of our model. 
The F-terms take at the minimum the expectation values
\beqa 
&|F^S|^2 = 8 e^K (S + \bar{S})^2 a^{2}(\pi \rho) |W_2|^2&\nn 
&\times (1-a^{2}(\pi \rho ))^2 \neq 0 &\nn
&|F^T|^2 = 0,&
\eeqa
which means that supersymmetry is broken along the dilaton direction.
The potential energy at this vacuum is negative:
\bea
&V_{vac} = -\frac{8 |W_2|^2}{V_0 (M^3/k M^{2}_P)^3  M^{2}_P}
\frac{a^{2}(\pi \rho)}{1 - a^{2}(\pi \rho)}.& 
\eea
The mass of the canonically normalized radion is 
\bea
&m^{2}_R = \frac{24}{V_0 (M^3/k M^{2}_P)^3 } \frac{a^{2}(\pi \rho) |W_2|^2}
{(1 -a^{2}(\pi \rho) )} \frac{1}{M^{4}_P}&
\eea
and the gravitino mass term is given by the expression 
\bea
&m_{3/2} = \frac{2}{\sqrt{ V_0} (M^3/k M^{2}_P)^{3/2} } \frac{a(\pi \rho ) }{(1-a^{2}(\pi \rho ) )^{1/2}} \frac{|W_2|}{M^{2}_P}.&\nn
& &
\eea
It is interesting to compare these features to those of the models, 
which are low-energy limits of weakly and strongly coupled heterotic 
string theories. As is well known, in the leading, tree-level, approximation, 
heterotic string gives the four-dimensional supergravity which enjoys 
the no-scale structure \cite{cfkn}. The gauge kinetic functions in all,
visible and hidden, sectors, are universal and depend only on the 4d 
dilaton superfield $S$, $H=S, \; \pa_T H =0$. In addition, at the perturbative level there is no superpotential for moduli superfields $S$ and $T$. 
As a consequence, the effective potential is positive semi-definite
and takes the form $V=K_{S \bar{S}} |F^S|^2 \geq 0$. 
Hence, the vacuum configuration corresponds to $F^S =0$, but $F^T$, which 
doesn't enter the potential, is allowed to take a non-zero value, so that supersymmetry is broken along the $T$ direction in moduli space. The problem 
with this is that the scale of supersymmetry breakdown is arbitrary, and it 
can only be hoped that it becomes fixed at a proper value, after 
taking into account various perturbative and/or nonperturbative corrections  
to the Lagrangian. 
In the supergravity model which is the low energy limit of 
weakly coupled heterotic string with one-loop corrections, and at the same 
time 
that of a strongly coupled heterotic string where these corrections 
arise in classical expansion taking into account the presence of an 
extra dimension, see \cite{lln}, the situation changes. 
When the interplay of sources of supersymmetry breakdown, such as
condensates in various sectors and expectation value of the superpotential, is taken into 
account, it is possible to arrange for unbroken supersymmetry in the 
anti-deSitter background, or for vacua where both $F^S$ and $F^T$ are 
sizable - see \cite{elpt,lt} for details. In all these considerations the 
effects of the nonperturbative warping of an extra dimension, like the one in the RS models, were not taken into account. The model which was constructed in \cite{flp3} finally allows to discuss the impact of the rapidly changing
warp-factor on the low-energy physics. In the model with just expectation 
values of the superpotentials serving as a supersymmetry breaking source, 
it is crucial that the effective superpotential acquires the exponential 
dependence on the modulus $T$. This leads to vanishing of the $F^T$ upon 
using the equation of motion for $T$. As a result, this time it is $F^S$ 
which is non-zero, although its value remains undetermined (as long as the 
 corrections to the Lagrangian are not taken into account). 
The vacuum energy takes the form
$V=-\frac{2}{M^{2}_P} e^K |W|^2 \leq 0$, which describes, at tree-level, 
an unstable background with negative energy density, in agreement with 
five-dimensional considerations.  
%no-scale models: 
%there $F^S =0$, $V_{vac} =0$, and $F^T$ is undetermined at tree-level 
%\cite{cfkn}.  

\section{Summary}

The main result of \cite{flp3} summarized in this talk is the 
four-dimensional 
effective supergravity action, which describes low-energy physics 
of the Randall--Sundrum model with moduli fields 
in the bulk and charged chiral matter living on the branes.

The low-energy
action has been read off from a compactification of a locally supersymmetric model in five dimensions. 
The exponential warp factor has interesting consequences for the form of the 
effective 4d supergravity. 
The asymmetry between the warped and unwarped walls 
is visible in the K\"ahler function, in the gauge kinetic functions and in the
superpotential. Roughly speaking the contributions to these functions
which come from the warped wall are suppressed by an exponential factor
containing the radion superfield. This is the way the warp factor and (and RS brane tensions)
is encoded in the low-energy Lagrangian. 

We have described the mechanism of supersymmetry breaking mediation which relies on a non-trivial configuration of the $Z_2$ odd  fields in the bulk. We point out,
that the odd $Z_2$ parity fields can be an important ingredient of 5d supersymmetric models. They play a
crucial role in communication between spatially separated branes. 

After freezing the dilaton, it is 
possible to stabilize the radion field in the backgrounds with broken
supersymmetry and excited odd-parity fields. 

We have shown that in the class of brane world models without charged matter 
and gauge fields in the bulk, unbroken $N=1$ supersymmetry implies 
vanishing cosmological constant. In the case where the sources that induce 
supersymmetry breakdown are represented simply by constant superpotentials on 
the branes, broken supersymmetry gives rise to anti-deSitter-type geometry in 
four dimensions. Hopefully, in more sophisticated models with various sources 
of supersymmetry breaking participating in the game, we shall be able to find 
models with broken supersymmetry and vanishing vacuum energy.  

We believe that the class of models we have constructed in \cite{flp3}  
provides a useful, explicit, setup to study low-energy phenomenology of 
the supersymmetric brane models with warped vacua.

\end{document}